\documentstyle[prb,aps]{revtex}
\begin{document}
\draft
\title{Backflow and dissipation during the quantum decay
of a metastable Fermi liquid}
\author{Kei Iida}
\address{Department of Physics, University of Tokyo, 
7-3-1 Hongo, Bunkyo, Tokyo 113-0033, Japan}
\date{\today}
\maketitle
\begin{abstract}
The particle current in a metastable Fermi liquid against a 
first-order phase transition is calculated at zero temperature.  
During fluctuations of a droplet of the stable phase, 
in accordance with the conservation law, not only does
an unperturbed current arise from the continuity at the boundary, 
but a backflow is induced by the density response.  
Quasiparticles carrying these currents are scattered by 
the boundary, yielding a dissipative backflow around the droplet.
An energy of the hydrodynamic mass flow of the liquid 
and a friction force exerted on the droplet by the quasiparticles
have been obtained in terms of a potential of their interaction 
with the droplet.

\end{abstract}

\section{Introduction}

A first-order phase transition at a temperature close to zero 
is thought to proceed via quantum tunneling through an energy 
barrier between the metastable and stable states in the 
configuration space.  Such a quantum decay of a metastable 
Fermi-liquid phase has attracted considerable attention
since it was found\cite{Satoh} in recent experiments of 
the phase separation from supersaturated $^{3}$He--$^{4}$He 
liquid mixtures that the observed degrees of critical supersaturation
were roughly
independent of temperatures below about 10 mK\@.
Another possible example is a deconfinement phase transition 
in cold, dense nuclear matter as encountered in a 
neutron star;\cite{Iida} this may take place during compression 
of the nuclear matter, induced for instance by mass accretion 
onto a neutron star from the companion in a close binary system.
As long as a homogeneous medium whose pressure is near 
the phase-equilibrium point adjusts adiabatically to fluctuations 
of a virtual droplet of the stable phase, the rate of 
such a quantum nucleation may be calculated\cite{LK} 
by treating the droplet radius $R$ as a macroscopic variable, 
deriving its Lagrangian from thermodynamic quantities of the system,
and describing the tunneling motion of $R$ in a semiclassical,
reversible fashion.

In a real system, however, such an adjustment proceeds 
at a finite rate.  Consequently, fluctuations of a virtual droplet 
are accompanied by elementary excitations in the metastable 
Fermi liquid.  These excitations not only renormalize the mass and 
the potential for $R$, but also lead to energy dissipation. 
Such renormalization and dissipation act to 
alter the tunneling probability exponentially.\cite{CL} 
In a hydrodynamic regime where the mean free path $l$ of
the excitations is much shorter than $R$, the particle current
is completely determined by the continuity equation, and
the energy dissipation is described in terms of the kinematic 
viscosity of the liquid.\cite{BD}
When the temperature is lowered enough to turn
the system into a collisionless regime ($R\ll l$), 
the low-lying excitations appear as quasiparticles. 
They are scattered by the droplet, and the resulting density 
fluctuation gives rise to a friction force exerted on the droplet 
as well as a dissipative backflow current.\cite{BD} 
Here we shall investigate how the flow pattern and the energy 
dissipation depend on coupling of the quasiparticles 
with the droplet.

Backflow around a slowly moving massive impurity in a Fermi 
liquid has been recently calculated by Zwerger;\cite{Z}
the result has been expressed in terms of phase shifts for 
scattering off the impurity.  Within linear response, 
the backflow is dipolar and proportional to the density 
response function.\cite{PN}  At long distances, however, 
the leading term is not dipolar, but is the radial contribution 
of at least second order in the scattering potential.
It was noted that this radial backflow arises from
the asymmetry in density around the fixed scatterer, 
induced by a finite background current.\cite{L} 
In this paper, we extend these calculations by Zwerger\cite{Z} 
to a spherically symmetric backflow, surrounding a 
fluctuational droplet of the stable phase, in a noninteracting, 
neutral Fermi gas at zero temperature.
We thus find that
the backflow current induced by the linear response has 
the same radial dependence as an unperturbed current,
and that these currents are governed by the particles 
on the Fermi surface.  The friction and the dissipative backflow
arising from scattering of these particles off the boundary
are expressed in terms of the transport cross sections.
Using the backflow pattern and friction so expressed, 
we derive the kinetic energy and the rate of energy dissipation
of the metastable gas.  Generalization to the interacting 
case is made within the Landau theory of Fermi liquids.
Finally, the backflow and dissipation in a charged Fermi liquid
relevant to dense nuclear matter are examined. 

\section{Particle Current and Dissipation in Noninteracting Systems}

We first consider a case in which a noninteracting, 
three-dimensional Fermi gas, composed of a single component 
(spin 1/2, bare mass $m$, and electric charge 0) and having 
an equilibrium number density $n_{1}$, is in a metastable state 
against a first-order phase transition at zero temperature.
In this case, a spherical droplet of the stable phase,
which is assumed to be incompressible and hence to be of
uniform number density $n_{2}$, develops 
inside, under, or outside the potential barrier.   
Such development, for pressures so close to the 
phase-equilibrium point that the droplet size is macroscopic, 
can be reduced to fluctuations of the droplet radius $R(t)$.
Let us now assume that $|{\dot R}|$ is far smaller than the
velocity of sound $c_{s}$ in the metastable Fermi gas,
i.e., the gas is nearly uniform.
This also means that $|{\dot R}|\ll v_{F}$, where 
$v_{F}=\hbar k_{F}/m$ with $k_{F}=(3\pi^{2}n_{1})^{1/3}$ 
is the Fermi velocity.
In the absence of the interaction of the droplet with the medium,
the droplet surface acts only as a boundary condition for 
the unperturbed current density ${\bf j}(r,t)|_{0}$: 
$-n_{2}{\dot R}{\hat{\bf r}}={\bf j}(R,t)|_{0}
-n_{1}{\dot R}{\hat{\bf r}}$.  
Here $r$ and ${\hat{\bf r}}$ are the length and the unit vector
of the position vector ${\bf r}$, of which the origin is 
set to be the droplet center. 
Under such a condition, the conservation law yields
\begin{equation}
{\bf j}(r,t)|_{0}=\left\{\begin{array}{lll}
  (n_{1}-n_{2}){\dot R}
  \left(\displaystyle{\frac{R}{r}}\right)^{2}{\hat{\bf r}}\ ,
         & \mbox{$r\geq R$} \\
             \\
         0\ ,
         & \mbox{$r<R\ .$}
 \end{array} \right.
\end{equation}

The actual current density ${\bf j}({\bf r},t)$ deviates from 
${\bf j}(r,t)|_{0}$ by the backflow due to the interaction 
of particles in the medium with the droplet. 
We describe this interaction in terms of a spherically 
symmetric, quasistatic potential $V(r,t)=\int d{\bf r}'
n_{\rm ex}({\bf r}',t)v(|{\bf r}-{\bf r}'|)$, where 
$n_{\rm ex}({\bf r},t)=n_{2}\theta[R(t)-r]$ is the density 
perturbation, and $v(r)$ is the potential of interaction of
a particle in the medium with that in the droplet.
Within linear response in $v(r)$, the backflow is
induced by $n_{\rm ex}$, the Fourier transform of which is given by
\begin{eqnarray*}
n_{\rm ex}({\bf q},\omega)&=&2\pi n_{2}\delta(\omega)
4\pi R^{3}\left[\frac{\sin qR}{(qR)^{3}}
               -\frac{\cos qR}{(qR)^{2}}\right]
\end{eqnarray*}
in the limit ${\dot R}\rightarrow 0$.
The response is characterized by the density fluctuation,\cite{PN}
\begin{equation}
n({\bf q},\omega)|_{\rm res}=v(q)\chi({\bf q},\omega)
n_{\rm ex}({\bf q},\omega)\ ,
\end{equation}
where $v(q)$ is the Fourier transform of $v(r)$, and
$\chi({\bf q},\omega)$ is the density-response function.
Let us assume 
that the range $a$ of the interaction $v$ is short, i.e., $a\ll R$,
and set $v(q)$ as $v(q=0)/(1+a^{2}q^{2})$ for simplicity.
Then, 
the macroscopic change in the number of particles near the droplet
may be calculated from Eq.\ (2) up to $O(a/R)$ as 
\begin{equation}
\Delta N|_{\rm res}=-2\pi 
\frac{\partial n_{1}}{\partial \mu_{1}}n_{2}v(q=0)aR^{2}\ ,
\end{equation}
where the asymptotic behavior of lim$_{q\rightarrow 0}$
$\chi({\bf q},0)=-\partial n_{1}/\partial \mu_{1}$ with the
chemical potential of the gas $\mu_{1}=\hbar^{2}k_{F}^{2}/2m$ is used.
By substituting Eq.\ (3)
into the conservation law, the radial backflow at large distances 
is obtained as
\begin{equation}
{\bf j}(r,t)|_{\rm res}=
-\frac{\partial n_{1}}{\partial \mu_{1}}
\frac{n_{2}}{n_{2}-n_{1}}v(q=0)\frac{a}{R}
{\bf j}(r,t)|_{0}\ .
\end{equation}
The proportionality of ${\bf j}|_{\rm res}$ to $aR{\dot R}$ reflects 
the fact that the change $\Delta N|_{\rm res}$
in the number of particles due to the
linear response behaves as $aR^{2}$.
According to whether the interaction is repulsive or attractive, 
the backflow current ${\bf j}|_{\rm res}$ goes in the same or 
the opposite direction with respect to ${\dot R}{\hat{\bf r}}$.
Note that Friedel-type oscillations appearing
near the droplet are not influential 
as long as $R$ is macroscopic, i.e., $k_{F}R\gg1$. 
Then, 
the backflow of form (4) is expected to approximately express 
the flow pattern for $r>R$.
The kinetic energy $K$ of the metastable Fermi gas can thus be 
calculated from the current sum ${\bf j}_{\rm mac}
={\bf j}|_{0}+{\bf j}|_{\rm res}$ as
\begin{eqnarray} 
K&=&\frac{1}{2}\int d{\bf r} m\frac{|{\bf j}_{\rm mac}|^{2}}{n_{1}}
\nonumber \\ &=&
2\pi mn_{1}R^{3}{\dot R}^{2}\left(1-\frac{n_{2}}{n_{1}}\right)^{2}
\left[1-2\frac{\partial n_{1}}{\partial\mu_{1}}
\frac{n_{2}}{n_{2}-n_{1}}v(q=0)\frac{a}{R}
+O(a^{2}/R^{2})\right]\ .
\end{eqnarray}
It is important to note that 
$K$ is directly related\cite{LK} to the reversible motion 
of $R$, since ${\bf j}_{\rm mac}$ satisfies the conservation law
on a macroscopic time scale of $R/|{\dot R}|$.
In the incompressible limit, Eq.\ (5) reduces to the result of
Lifshitz and Kagan.\cite{LK}

The macroscopic current ${\bf j}_{\rm mac}$ in turn 
causes a backflow current beyond linear response through
the scattering of particles in the medium off the droplet.  
Let us assume the droplet to be inert and
characterize the behavior of the particles 
by the incoming momenta ${\bf k}$ and the outgoing scattering states
in $V$: $\psi_{k}({\bf r})=\langle{\bf r}|{\bf k}+\rangle$. 
By noting the correspondence with a local reference frame 
moving with the velocity 
${\bf v}_{\rm mac}={\bf j}_{\rm mac}/n_{1}$, 
the equilibrium distribution of the particles
is obtained as a shifted Fermi distribution function, 
$n^{0}(\varepsilon_{{\bf k}-\frac{m}{\hbar}{\bf v}_{\rm mac}})$, 
with $\varepsilon_{\bf k}=\hbar^{2}k^{2}/2m$.
The actual current density may thus be calculated up to
linear order in ${\dot R}$ as
\begin{eqnarray}
{\bf j}({\bf r},t) &=& \frac{2}{(2\pi)^{3}}\int d{\bf k}
\frac{\hbar}{m}
{\rm Im}\left[\psi_{k}^{*}({\bf r})\frac{\partial}{\partial {\bf r}}
        \psi_{k}({\bf r})\right] 
n^{0}(\varepsilon_{{\bf k}-\frac{m}{\hbar}{\bf v}_{\rm mac}})
 \nonumber \\ &=&
\frac{2k_{F}^{2}}{(2\pi)^{3}}\int d\Omega_{k}
{\bf v}_{\rm mac}\cdot {\bf{\hat k}}
{\rm Im}\left.\left[\psi_{k}^{*}({\bf r})\frac{\partial}
        {\partial {\bf r}}\psi_{k}({\bf r})
        \right]\right|_{k=k_{F}}\ ,
\end{eqnarray}
where ${\bf{\hat k}}$ is the unit vector of ${\bf k}$, and
$d\Omega_{k}$ is the integration over the directions of 
${\bf{\hat k}}$.  Equation (6) shows that the flow is 
governed by the particles on the Fermi surface.
At large distances, by using such an asymptotic form of 
$\psi_{k}({\bf r})$ as $e^{i{\bf k}\cdot{\bf r}}+
f_{k}({\bf{\hat k}}\cdot{\bf{\hat r}})e^{ikr}/r+\ldots$ with
the usual scattering amplitude $f_{k}$, we obtain the backflow current 
$\delta{\bf j}|_{\rm curr}={\bf j}-{\bf j}_{\rm mac}$ as
\begin{eqnarray}
\delta{\bf j}|_{\rm curr}&=&
\frac{2k_{F}^{3}}{(2\pi)^{3}}\int d\Omega_{k}
{\bf v}_{\rm mac}\cdot {\bf{\hat k}}
\left\{\frac{|f_{k}|^{2}}{r^{2}}{\bf{\hat r}}+
{\rm Im}\left[ie^{i(kr-{\bf k}\cdot{\bf r})}
\frac{f_{k}}{r}{\bf{\hat r}}+ie^{i({\bf k}\cdot{\bf r}-kr)}
\frac{f^{*}_{k}}{r}{\bf{\hat k}}
\right]\right\}_{k=k_{F}}+O(r^{-3})
\nonumber \\ &=&
 -\frac{3\sigma_{\rm tr}(k_{F})}{4\pi r^{2}}
                           {\bf j}_{\rm mac}+O(r^{-3})\ ,
\end{eqnarray}
where $\sigma_{\rm tr}(k)=\int d\Omega_{k}
(1-{\bf{\hat k}}\cdot{\bf{\hat r}})|f_{k}|^{2}$ 
is the exact scattering cross section.
Here we have used the asymptotic form 
$e^{i(kr-{\bf k}\cdot{\bf r})}\rightarrow 
i\delta(\Omega_{k}-\Omega_{r})/2kr$
as well as the optical theorem. 
We note that ${\bf j}_{\rm mac}$ comes from Eq.\ (6) 
via the unperturbed term $e^{i{\bf k}\cdot{\bf r}}$.
We see from Eq.\ (7) that the backflow 
$\delta{\bf j}|_{\rm curr}$ is at least of order $V^{2}$ and 
is induced by the macroscopic current ${\bf j}_{\rm mac}$
in its opposite direction.

The physical origin of the backflow $\delta{\bf j}|_{\rm curr}$
is analogous to that given by Landauer\cite{L} 
for a backflow around a slowly moving impurity. 
A portion of the metastable Fermi gas, which carries
the macroscopic current ${\bf j}_{\rm mac}$
but cannot adjust adiabatically to the development of the droplet,
is scattered by the boundary, leading to a density fluctuation
around the droplet.  This fluctuation in turn diffuses 
at a velocity of $v_{F}$, yielding the asymptotic backflow
current written by Eq.\ (7). 
This current has nonvanishing divergence, since 
the density fluctuation is positive or negative definite.
This is a contrast to the case of a slowly moving impurity
in which the density fluctuation, being positive in front of
and negative behind the scatterer, is distributed
in such a way that the corresponding backflow has 
vanishing divergence.\cite{Z}

We proceed to calculate the friction force ${\bf F}$ due to 
momentum transfer from the medium to the boundary.
This force may be expressed as\cite{Z}
\begin{equation}
{\bf F}=\frac{2}{(2\pi)^{3}}\int d{\bf k}
\left\langle{\bf k}+\left|\frac{\partial V}{\partial{\bf r}}
\right|{\bf k}+\right\rangle
n^{0}(\varepsilon_{{\bf k}-\frac{m}{\hbar}
{\bf v}_{\rm mac}|_{r=R}})\ .
\end{equation}
${\bf F}$ originates from the density fluctuation
induced by the macroscopic current ${\bf j}_{\rm mac}$, 
since it vanishes when ${\bf v}_{\rm mac}|_{r=R}=0$. 
Up to $O({\dot R})$ and $O(a/R)$,  we obtain
\begin{eqnarray}
{\bf F}&=&\frac{2}{(2\pi)^{3}} \int d\Omega_{k}
\frac{mk_{F}^{2}}{\hbar}{\bf v}_{\rm mac}|_{r=R}
\cdot{\bf{\hat k}}
\left\langle{\bf k}+\left|\frac{\partial V}{\partial{\bf r}}
\right|{\bf k}+\right\rangle|_{k=k_{F}}
\nonumber \\ &=&
\left[1-\frac{\partial n_{1}}{\partial \mu_{1}}
\frac{n_{2}}{n_{2}-n_{1}}v(q=0)\frac{a}{R}\right]
n_{1}\hbar k_{F}\sigma_{\rm tr}(k_{F})
\left(1-\frac{n_{2}}{n_{1}}\right){\dot R}
{\bf{\hat r}}\ .
\end{eqnarray}
Here we have used the relation\cite{BS}
$\langle{\bf k}+|\partial V/\partial{\bf r}|
{\bf k}+\rangle=(\hbar^{2}k^{2}/m)\sigma_{\rm tr}(k){\bf{\hat k}}$, 
derived from the Lippmann-Schwinger equation.
The rate of energy dissipation due to this friction may
then be calculated as 
${\dot E}=-4\pi{\bf F}\cdot{\bf v}_{\rm mac}|_{r=R}$.
From Eqs.\ (1), (4), and (9), we obtain
\begin{equation}
{\dot E}=-4\pi^{2}\alpha 
\left[1-2\frac{\partial n_{1}}{\partial \mu_{1}}
\frac{n_{2}}{n_{2}-n_{1}}v(q=0)\frac{a}{R}+O(a^{2}/R^{2})\right]
n_{1}\hbar k_{F}
\left(1-\frac{n_{2}}{n_{1}}\right)^{2}R^{2}{\dot R}^{2}\ ,
\end{equation}
with $\alpha=\sigma_{\rm tr}(k_{F})/\pi R^{2}$.
We thus find that Eq.\ (10) is similar to the result obtained
by Burmistrov and Dubovskii\cite{BD} using a dimensional analysis
based on the expression for ${\dot E}$ in the hydrodynamic regime.
Their result, however, has left unknown the coefficient of
$n_{1}\hbar k_{F}(1-n_{2}/n_{1})^{2}R^{2}{\dot R}^{2}$,
which is now expressed in terms of the potential $v$.

\section{Extension to Interacting and Charged Systems}

Short-range interaction $\tilde{v}$ between particles in the medium 
plays a role in modifying the backflow current and the rate 
of energy dissipation calculated for a noninteracting Fermi gas.
This role can be examined within the Landau theory
of Fermi liquids,\cite{PN} which allows one to construct
a quasiparticle state $\tilde{\psi}_{p}$ with momentum ${\bf p}$
from the corresponding noninteracting state
via adiabatic switching on of the interaction $\tilde{v}$.
In order to obtain the distribution 
$n_{\bf p}$ of the quasiparticles,
it is sufficient to notice that a transport equation 
for the quasiparticles is identical with 
the corresponding equation for a noninteracting system,
except that the velocity term
is determined by the quasiparticle energy 
$\tilde{\varepsilon}_{\bf p}$ and that the force
$-\partial\tilde{\varepsilon}_{\bf p}/\partial{\bf r}$, 
exerted by other quasiparticles, emerges.
The transport equation thus reads
\begin{equation}
\frac{\partial n_{\bf p}}{\partial t}
+\frac{\partial n_{\bf p}}{\partial{\bf r}}
 \cdot\frac{\partial\tilde{\varepsilon}_{\bf p}}{\partial{\bf p}}
-\frac{\partial n_{\bf p}}{\partial{\bf p}}
 \cdot\frac{\partial(\tilde{\varepsilon}_{\bf p}+V)}
           {\partial{\bf r}}=0\ .
\end{equation}
Let us divide $n_{\bf p}$ into the equilibrium part
$\langle n_{\bf p}\rangle$ and the fluctuating part 
$\delta n_{\bf p}$.
For the translationally invariant system considered here and
$\partial n_{\bf p}/\partial t=0$, we obtain
$\langle n_{\bf p}\rangle
=n^{0}(\varepsilon^{0}_{\frac{1}{\hbar}{\bf p}
-\frac{m}{\hbar}{\bf v}_{\rm mac}})$,
where $\varepsilon^{0}_{\bf k}=\hbar^{2}k^{2}/2m^{*}$ 
with the quasiparticle effective mass $m^{*}$, and
$\delta n_{\bf p}=
\langle n_{\bf p}\rangle(|\tilde{\psi}_{p}|^{2}-1)$.
Note that $\langle n_{\bf p}\rangle$ and $\delta n_{\bf p}$,
associated with the current densities ${\bf j}_{\rm mac}$ and 
$\delta{\bf j}|_{\rm curr}$, reduce to the similar expressions
$n^{0}(\varepsilon_{\frac{1}{\hbar}{\bf p}
-\frac{m}{\hbar}{\bf v}_{\rm mac}})$ and 
$n^{0}(\varepsilon_{\frac{1}{\hbar}{\bf p}
-\frac{m}{\hbar}{\bf v}_{\rm mac}})
(|\psi_{p/\hbar}|^{2}-1)$ for $\tilde{v}=0$. 
Consequently, ${\bf j}|_{0}$,
${\bf j}|_{\rm res}$, and $\delta{\bf j}|_{\rm curr}$
are still described by Eqs.\ (1), (4), and (7)
in which the scattering amplitude $f_{k}$
and the thermodynamic quantity $\partial\mu_{1}/\partial n_{1}$ 
are modified by the interaction $\tilde{v}$. 
Here, the interaction potential $v$
is assumed to be unchanged by $\tilde{v}$. 
This holds if the potential $v$ is a scalar field.\cite{PN}
By using the obtained current densities and 
distribution of the quasiparticles, we can calculate 
the kinetic energy $K$, the friction force ${\bf F}$,
and the rate of energy dissipation ${\dot E}$. 
The results are given by Eqs.\ (5), (9), and (10) in which
the scattering amplitude $f_{k}$
and the thermodynamic quantity $\partial\mu_{1}/\partial n_{1}$ 
are those calculated for an interacting system,
and the momentum of unit volume $n_{1}\hbar k_{F}$ is multiplied
by $m/m^{*}$.
Note that the generalization to the interacting case 
mentioned above is valid for a 
Fermi liquid at the low temperatures that ensure $l\gg R$.\cite{note}
At higher temperatures, where quasiparticle collisions should be 
built into the transport equation as a collision integral, 
the current and the rate of energy dissipation
remain to be examined.

Next, with reference to dense nuclear matter that is neutralized 
by a gas of electrons and muons and is metastable against 
deconfinement,\cite{Iida}  we consider a case in which 
a metastable Fermi liquid consists of particles of charge $q_{1}e$
that interact mainly via the short-range potential $\tilde{v}$ 
and are embedded in a uniform neutralizing background, 
and a stable phase is immersed in the same background. Then,
a droplet of the stable phase has nonvanishing charge 
as long as $n_{1}\neq n_{2}$. 
This charge is in turn screened by the charged particles
in the medium. The resulting particle migration
is characterized by the Thomas-Fermi screening length
$\lambda_{\rm TF}=[(\partial\mu_{1}/\partial n_{1})
/4\pi(q_{1}e)^{2}]^{1/2}$, since $|{\dot R}|\ll c_{s}$, 
being assumed in the medium, is consistent with the condition 
$\lambda_{\rm TF}\gg R$, appropriate for the linear Thomas-Fermi
analysis\cite{D} of the charge response of the medium.
Let us take account of
the electrostatic potential $\phi$ of interaction 
between the particles inside and outside the droplet
and the charge density perturbation 
$\rho_{\rm ex}=q_{1}e(n_{2}-n_{1})\theta[R(t)-r]$,
in addition to the short-range potential $v$ and
the density perturbation $n_{\rm ex}$.  
Accordingly, the Coulomb-induced backflow ${\bf j}|_{\rm res}^{(C)}$
of linear order arises along with the backflow 
${\bf j}|_{\rm res}$ given by Eq.\ (4).
When $r\gg\lambda_{\rm TF}$ is satisfied,
one obtains ${\bf j}|_{\rm res}^{(C)}=-{\bf j}|_{0}$ 
since perfect screening ensures cancellation 
between the unperturbed current ${\bf j}|_{0}$ and 
the backflow current ${\bf j}|_{\rm res}^{(C)}$.
At distances $r\lesssim\lambda_{\rm TF}$, however,
the screening action of the particles gives rise to 
slight inhomogeneity of order $(R/\lambda_{\rm TF})^{2}$
in the density distribution of the medium, which we have
calculated within the linear Thomas-Fermi approximation.\cite{IS}
The resulting density distribution yields the backflow 
${\bf j}|_{\rm res}^{(C)}$ via the continuity equation:
\begin{equation}
{\bf j}|_{\rm res}^{(C)}=\{-1+[1-\kappa R+O(\kappa^{2}R^{2})]
(1+\kappa r)e^{-\kappa(r-R)}\}{\bf j}|_{0}\ ,
\end{equation}
with $\kappa=\lambda_{\rm TF}^{-1}$. 
Up to $O(\kappa R)$ and $O(a/R)$, the kinetic energy $K$, 
the dissipative backflow $\delta{\bf j}|_{\rm curr}$,
the friction force ${\bf F}$, and the dissipation rate ${\dot E}$
can be calculated by following a line of argument for 
a neutral Fermi liquid and by incorporating Eq.\ (12) 
in the macroscopic current ${\bf j}_{\rm mac}$
and modifications by $\phi$ in the scattering amplitude $f_{k}$. 
Since ${\bf j}|_{\rm res}^{(C)}\sim O(\kappa^{2}R^{2})$ at $r=R$, 
the resultant expressions for ${\bf F}$ and ${\dot E}$ are 
identical with Eqs.\ (9) and (10), respectively, except that 
$n_{1}\hbar k_{F}$ is replaced by $(m/m^{*})n_{1}\hbar k_{F}$. 
On the other hand, $K$ and $\delta{\bf j}|_{\rm curr}$ have
corrections due to ${\bf j}|_{\rm res}^{(C)}$; the expression for
$\delta{\bf j}|_{\rm curr}$ is given by Eq.\ (7) in which 
${\bf j}_{\rm mac}={\bf j}|_{0}+{\bf j}|_{\rm res}+
{\bf j}|_{\rm res}^{(C)}$, and that for $K$ is obtained as
\begin{equation} 
K=2\pi mn_{1}R^{3}{\dot R}^{2}
\left(1-\frac{n_{2}}{n_{1}}\right)^{2}
\left[1-2\frac{\partial n_{1}}{\partial\mu_{1}}
\frac{n_{2}}{n_{2}-n_{1}}v(q=0)\frac{a}{R}
-\frac{3}{2}\kappa R\right]\ .
\end{equation}
In Eq.\ (13), the term of order $\kappa R$ reflects the role 
played by ${\bf j}|_{\rm res}^{(C)}$ 
in cancelling the unperturbed current ${\bf j}|_{0}$ at
$r\gtrsim\lambda_{\rm TF}$.

\section{Conclusion}

We have evaluated the flow pattern and 
the friction induced during nucleation of the stable phase
in a metastable Fermi liquid.  The kinetic energy $K$ and 
the dissipation rate ${\dot E}$ of the liquid,
which are relevant to the calculations of the rate of 
quantum nucleation, have been expressed in terms of 
the interaction potential $v$.
In order to make practical estimates,
one needs information as to $v$,
which
is generally hard to obtain from experiments.
The corresponding scattering potential $V$
may have an inelastic property
if the droplet undergoes excitations due to the
scattering of quasiparticles in the medium off the boundary.
It is important to recall that the results for the current 
and dissipation as obtained above are based on 
the assumption of the liquid being nearly incompressible.
The effect of compressibility not only develops 
the nonlinear response contribution to the macroscopic current 
${\bf j}_{\rm mac}$,  but also yields a reduction in the 
kinetic energy $K$ via the emission of sound.
The latter, which stems from the finite $\omega$ behavior of 
${\bf j}_{\rm mac}$, plays a role in increasing the nucleation 
rate exponentially.\cite{K}

Finally, we consider a case in which 
the liquid is in a superfluid state.
The superfluidity has consequence in reducing the available 
quasiparticle states with momenta close to the Fermi surface
and hence in weakening the friction force ${\bf F}$.
This effect may be estimated 
by changing, in the Fermi distribution in Eq.\ (8),
the shifted quasiparticle energy 
$\varepsilon^{0}_{{\bf k}-\frac{m}{\hbar}{\bf v}_{\rm mac}}$
into the quantity $\sqrt{(\varepsilon^{0}_{\bf k}
-\hbar^{2}k_{F}^{2}/2m^{*})^{2}+
\Delta_{\bf k}^{2}}-(m/m^{*})\hbar{\bf k}\cdot{\bf v}_{\rm mac}
+\hbar^{2}k_{F}^{2}/2m^{*}$, 
with the pairing gap $\Delta_{\bf k}$.
This indicates that, when the macroscopic current flows  
slowly compared with the critical velocity that makes 
$\Delta_{\bf k}$ vanish, i.e., 
$|{\bf v}_{\rm mac}|<(m^{*}/m)\Delta_{\bf k}/\hbar k$,  
and the temperature is zero, the friction does not occur.
This is because
Cooper pairs sustain ${\bf j}_{\rm mac}$ as a supercurrent.
With increasing temperature and/or flow velocity,
however, quasiparticle excitations become prevailing,
giving rise to the dissipation of energy.
We thus find that the nucleation rate
depends strongly on the temperature and
the flow velocity, a feature expected to be 
ascertained by future helium experiments at ultralow
temperatures.
  
\section*{Acknowledgments}

The author thanks S. Ichimaru for critical reading 
of the original manuscript, and S. N. Burmistrov and 
T. Satoh for discussion.
He acknowledges the hospitality of Aspen Center for Physics,
where this work was completed.
This work was supported in part by Grant-in-Aid for
Scientific Research provided by the Ministry of Education,
Science, and Culture of Japan through Grants No.\ 07CE2002
and No.\ 3687.


\begin{references}
\bibitem{note} At nonzero temperatures, for slow migration 
of the droplet surface as considered here, 
the backflow $\delta{\bf j}|_{\rm curr}$ at $r\gtrsim l$ 
should acquire a component that depends on $l/r$.  
This component, whose calculations are beyond the scope of this 
paper, has no effect on the migration itself as long as $l\gg R$.  
In this case, such a migration, mainly determined by the particle 
current in the neighborhood of the droplet, is independent of the 
quasiparticle collisions.
\end{references}
\end{document}